\documentclass[prd,twocolumn,showpacs,amsmath,amssymb]{revtex4}
\usepackage{graphicx}
\usepackage{dcolumn}
\usepackage{bm}

\begin{document}

\title{Warm inflation in loop quantum cosmology: a model with a general dissipative coefficient}

\author{Xiao-Min Zhang}
\email{zhangxm@mail.bnu.edu.cn}
\affiliation{Department of Physics, Beijing Normal University,
Beijing 100875,  People's Republic of China}
\author{Jian-Yang Zhu}
\thanks{Corresponding author}
\email{zhujy@bnu.edu.cn}
\affiliation{Department of Physics, Beijing Normal University,
Beijing 100875,  People's Republic of China}

\begin{abstract}
A general form of warm inflation with the dissipative coefficient  $\Gamma=\Gamma _0\left( \phi /\phi
_0\right) ^n\left( T/\tau _0\right) ^m$ in loop quantum cosmology is studied. In this case, we obtain conditions for the existence of a warm inflationary attractor in the context of loop quantum cosmology by using the method of stability analysis. The two cases when the dissipative coefficient is independent $(m=0)$ and dependent $(m\neq0)$ on temperature are analyzed specifically. In the latter case, we use the new power spectrum which should be used when considering temperature dependence in the dissipative coefficient. We find that the thermal effect is enhanced in the case $m>0$. As in the standard inflation in loop quantum cosmology, we also reach the conclusion that quantum effect leaves a tiny imprint on the CMB sky.
\end{abstract}

\pacs{98.80.Cq, 98.80.Qc}
\maketitle

\section{\label{sec:level1}Introduction}

Inflation, as a necessary supplement to the standard cosmological model, can
successfully explain many long-standing problems such as horizon and
flatness. Inflation is also a good mechanism to explain the large scale
features of the universe \cite{Guth,Linde,Liddle2000}. The standard
inflation needs a reheating period to end the inflation and start the
radiation dominated phase. Another type of inflation called ``warm
inflation'' was proposed by Berera and Fang \cite{BereraFang}. From then
many works have been done in the context of warm inflation, such as tachyon
warm inflationary model \cite{tachyonWI} and natural warm inflation \cite
{NWI}, but most of them are focused on a classical universe case. In the
scenario of warm inflation, the inflation of the universe is accompanied by
continuous radiation producing, so the universe is hot during inflation and
can go into radiation dominated phase smoothly. In order to realize the
acceleration of the universe, the dominant energy is still the potential of
the inflaton. The radiation producing is due to the interaction between the
inflaton and other fields. After considering the dissipative interaction, a
dissipative term is contained in the equation of motion of the inflaton.
Different dissipative coefficient (denoted as $\Gamma $) can be obtained
when considering different microscopic models. We will use a most general
form of dissipative coefficient $\Gamma =\Gamma _0\left( \phi /\phi
_0\right) ^n\left( T/\tau _0\right) ^m$ by concluding all the cases and try
to give a general picture. Besides these, warm inflation can also eliminate
the ``$\eta $-problem'' \cite{etaproblem} and decrease the overlarge
amplitude of inflaton in standard inflation \cite{Berera2005}. A most
discriminative feature of the warm inflation is that the main contribution to
the density perturbation is the thermal fluctuations rather than vacuum
fluctuation \cite{BereraFang,Lisa2004,Berera2000}. The power spectrum in
warm inflation is analyzed mostly in the case that the dissipative
coefficient is independent on temperature \cite{Lisa2004,Berera2000},
however, considering the microphysical basis of warm inflation in the very
early universe, the dissipative coefficient is dependent on temperature in
most cases and deserve more research. The new scalar perturbation spectra
for $\Gamma =\Gamma (\phi ,T)$ case is given by Chris Graham and Ian G. Moss
in \cite{ChrisIan}. We use the new scalar perturbation spectra when
analyzing the temperature dependent case.

On the other hand, loop quantum gravity (LQG) is a mathematically
well-defined, nonperturbative, and background independent quantization of
gravity \cite{LQG}. The space-time geometry in LQG is discrete when
approaching the Planck scale and become continuous in a large eigenvalue
limit. LQG is a good and pioneering scheme to unify quantum mechanics and
gravity. Loop quantum cosmology (LQC) is the application of LQG in the
homogeneous and isotropic universe \cite{LQC,LQC1,classicalH}. LQC inherits
quantization scheme and techniques from LQG but focuses on symmetry reduced
models and solves many longstanding problems. The singularity in general
relativity is replaced by a cosmological bounce in LQC. The underlying
dynamics of LQC is the discrete quantum difference equation of quantum
geometry which is not easy to solve. Fortunately, there are two approaches
to solve this problem, one is using sophisticated numerical methods \cite{PSingh},
and the other one is using semiclassical states to construct an effective theory of LQC by incorporating quantum
corrections to the classical dynamics. Generally speaking, there are two
kinds of quantum corrections, namely inverse-volume correction and holonomy
correction \cite{LQCcorrections}. In both cases, one can get a modified
Friedmann equation by making Hamilton analysis. The modified Friedmann
equation is very different from the classical one but can reduce to
classical dynamics when the quantum effect is quite weak. In this paper, we
use holonomy correction Friedmann equation to investigate warm inflation in
loop quantum cosmology (Warm-LQC). With holonomy correction, the Friedmann
equation acquires a quadratic density modification when the energy scale is
comparable to the critical energy density $\rho _c$ ($\rho _c\simeq 0.41\rho
_{pl}$, $\rho _{pl}=m_p^4$, $m_p^2=G^{-2}$) \cite{rhoc}. That the universe can undergo a
big bounce when the energy density equals the critical density can be seen
from the modified Friedmann equation. Cosmological inflation naturally
happens after the big bounce, and a new phase named super-inflation \cite{super-inflation} is found
before the normal inflation, which is a character of LQC. The standard
inflationary model in LQC was fully researched in Ref. \cite{lingyiJCAP,Singh2006}.
The warm inflationary model in LQC was also considered in Refs. \cite
{Herrera2010,Xiao2011}, which are a good try to combine warm inflation with LQC. But
for simplicity, they only discuss the case when the dissipative coefficient
is a constant. Thermal fluctuations were treated earlier in loop cosmology in \cite{Joao2007} in a noninflationary setting and they restricted themselves on the closed universe models. They assumed a purely thermal origin for the fluctuations and obtained a power spectrum through the partition function. However they can't get a nearly scale-invariant spectrum in classical universe. In order to get a nearly scale-invariant spectrum, they make use of inverse-volume correction in LQC and the problem in the classical unverse is
eliminated.

In this paper we try to give a general picture of a warm inflationary model in
LQC. We use the method of stability analysis to obtain the slow-roll
conditions in Warm-LQC when both the potential and the dissipative
coefficient might be dependent on temperature. We study two cases when the
dissipative coefficient is independent and dependent on temperature and give
a workable example in both cases. What we should note here is that in both cases we treat the perturbations
in Warm-LQC following the way in \cite{lingyiJCAP,Herrera2010}, for the quantum effect is not obvious when horizon crossing as we will argue later in the paper. More details and consistent works about perturbations in LQC is beyond the scope of our works and can be found in \cite{LQCperturbation,LQCperturbation1}. In the temperature dependent case, we take the new form
of the power spectrum which should be used when the dissipative coefficient
is temperature dependent. We use the  Wilkinson Microwave Anisotropy Probe (WMAP) data
to constrain parameters and detect the order of magnitudes of quantum effect.

The outline of the paper is as follows. In the next section we introduce the
effective theory of LQC. In Sec. \ref{sec:level3} the dynamics of Warm-LQC
is briefly introduced. We analyze the slow-roll conditions in Warm-LQC in
Sec.\ref{sec:level4}. The two cases when the dissipative coefficient is
independent and dependent on temperature are calculated respectively in Sec.%
\ref{sec:level5} and Sec.\ref{sec:level6}. Finally, we draw the conclusions
in Sec.\ref{sec:level7}.

\section{\label{sec:level2}Loop Quantum Cosmology}

In this section, we will introduce the effective theory of LQC based on
holonomy correction in the flat model of universe. In LQG, the phase space
of classical general relativity is expressed in terms of SU(2) connection $%
A_a^i$ and density-weighted triads $E_i^a$. After the symmetry reduction and
the gauge fixing from LQG, the only remaining degrees of freedom in the
phase space of LQC are the conjugate variables of connection $c$ and triad $p
$, which satisfy Poission bracket $\{c,p\}=\frac 13\gamma \kappa $, where $%
\kappa =8\pi G$, $\gamma $ is the Barbero-Immirzi parameter ($\gamma $ is
set to be $\gamma \simeq 0.2375$ by the black hole thermodynamics)\cite{BarberoImmirzi}.
In the Friedmann-Robertson-Walker (FRW)
cosmology, the two conjugate variables can be expressed as
\begin{equation}
c=\gamma \dot{a},\ p=a^2,  \label{cp}
\end{equation}
where $a$ is the FRW scale factor. The classical Hamiltonian constraint in
terms of the connection and triad is given by \cite{classicalH}
\begin{equation}
\mathcal{H}_{cl}=-\frac 3{\kappa \gamma ^2}\sqrt{p}c^2+\mathcal{H}_M,  \label{Hcl}
\end{equation}
where $\mathcal{H}_M$ is the matter Hamiltonian. Using holonomy correction to modify
the gravity part in the classical Hamiltonian constraint, the effective
Hamiltonian constraint is given by \cite{LQCcorrections}
\begin{equation}
\mathcal{H}_{eff}=-\frac 3{\kappa \gamma ^2\bar{\mu}^2}\sqrt{p}\sin ^2(\bar{\mu}c)+\mathcal{H}_M.
\label{Heff}
\end{equation}
where $\bar{\mu}$ corresponds to the dimensionless length of the edge of the
elementary loop over which the holonomies are computed, and the area is $%
{\cal A}=\bar{\mu}^2p=\alpha l_{pl}^2$ , where $\alpha $ is of order of
unity and $l_{pl}=\sqrt{\hbar G}$ is the Planck length. Using the
Hamiltonian constraint (\ref{Heff}) one can get the Hamiltonian equation for
$p$ :
\begin{equation}
\dot{p}=\{p,\mathcal{H}_{eff}\}=-\frac{\kappa \gamma }3\frac{\partial \mathcal{H}_{eff}}{%
\partial c}=\frac{2a}{\gamma \bar{\mu}}\sin (\bar{\mu}c)\cos (\bar{\mu}c),
\label{dotp}
\end{equation}
which combined with Eq. (\ref{cp}) implies that
\begin{equation}
\dot{a}=\frac 1{\gamma \bar{\mu}}\sin (\bar{\mu}c)\cos (\bar{\mu}c).
\label{dota}
\end{equation}
Furthermore, the vanishing Hamiltonian constraint $H_{eff}\approx 0$ yields
\begin{equation}
\sin ^2(\bar{\mu}c)=\frac{\kappa \gamma ^2\bar{\mu}^2}{3a}\mathcal{H}_M.
\label{sin2uc}
\end{equation}
From Eqs. (\ref{dota}) and (\ref{sin2uc}), the effective Friedmann equation
results in
\begin{equation}
H^2=\frac \kappa 3\rho \left( 1-\frac \rho {\rho _c}\right) ,
\label{effFriedmann}
\end{equation}
where $H=\dot{a}/a$ is the Hubble rate and the critical density $\rho _c$ is
given by
\begin{equation}
\rho _c=\frac 3{\kappa \gamma ^2\bar{\mu}^2a^2}=\frac{\sqrt{3}}{32\pi
^2\gamma ^3}\rho _{pl}\simeq0.41\rho_{pl},  \label{rhoc}
\end{equation}
where $\rho _{pl}=G^{-2}$ is the Planck density and $\alpha=4\sqrt{3}\pi\gamma$ \cite{rhoc} is used. Compared with the classical Friedmann equation, a $\rho ^2$ correction term is added to the effective
Friedmann in LQC, which implies the Hubble rate vanishes when $\rho =\rho _c$
and the universe undergoes a turnaround in the scale factor instead of
singularity. When $\rho \ll \rho _c$, the modification term is negligible
and the classical one is recovered.

\section{\label{sec:level3}Basic Equations of Warm-LQC}

We consider a spatially flat, homogeneous universe dominated by a scalar
field $\phi $ (inflaton) and radiation produced by the interaction of
inflaton with other fields which are subordinate. The interaction existing
during inflation seems more natural than the assumption in standard
inflation that the inflaton is an isolated, noninteracting field, so
instead of a steep supercooling phase in standard inflation, the universe
has a temperature $T$ during warm inflation. The scalar inflaton field must be
potential energy dominated to realize inflation. In the Warm-LQC scenario, the
evolution of the inflaton is governed by both the potential $V(\phi ,T)$ and
the dissipative coefficient $\Gamma (\phi ,T)$
\begin{equation}
\ddot{\phi}+(3H+\Gamma )\dot{\phi}+V_\phi =0  \label{EOMphi},
\end{equation}
where the subscripts denote a derivative. For simplicity, $\Gamma $ is often
set to be a constant in some papers \cite{Herrera2010,Xiao2011,Taylor2000}.
Considering some concrete model of the interaction between inflaton and
other fields, different forms of $\Gamma $ have been obtained \cite
{Ian0603266,BereraRamos}, for example in the supersymmetry (SUSY) low temperature case, $%
\Gamma \simeq 0.64\times g^2h^4(\frac{g\phi }{m_\chi })^4\frac{T^3}{(m_\chi
)^3}$ ($m_\chi \approx g\phi$), which was fully calculated in Ref. \cite{Mar2007}. Based on some
different forms of dissipative coefficient, a general form of dissipative
coefficient $\Gamma =C_\phi \frac{T^m}{\phi ^{m-1}}$ was proposed in \cite
{ZhangYi2009}. Here, we'll use a more general form of dissipative
coefficient which is the same as in Refs. \cite{Lisa2004,Campo2010}.
\begin{equation}
\Gamma =\Gamma _0\left( \frac \phi {\phi _0}\right) ^n\left( \frac T{\tau _0}%
\right) ^m,  \label{Gamma}
\end{equation}
with n and m integers and $\phi _0$, $\tau _0$, $\Gamma _0$ some
nonnegative constants. This kind of general form of dissipative coefficient
has the right dimension and can cover all different cases, for example, when
$m=n=0$, $\Gamma =\Gamma _0$ is recovered and when $m=3$, $n=-2$, the SUSY low temperature ($\Gamma
=C_\phi \frac{T^3}{\phi ^2})$ case is included.

 A parameter is defined to measure the damping strength of the Warm-LQC scenario
\begin{equation}
R=\frac \Gamma {3H}   \label{R}.
\end{equation}
 For strong dissipation
regime in Warm-LQC, we have $R\gg 1$, on the contrary, $R\ll 1$ is the weak dissipation
regime in Warm-LQC.

The dissipation of the inflaton's motion is associated with the production
of entropy. The expression for entropy density from thermodynamics is $%
s=-\partial f/\partial T$, when the free energy $f=\rho -Ts$ is dominated by
the potential, we have
\begin{equation}
s\simeq -V_T.  \label{s}
\end{equation}
The total energy density, including the contributions of the inflaton and
radiation, is
\begin{equation}
\rho =\frac 12\dot{\phi}^2+V(\phi ,T)+Ts.  \label{rho}
\end{equation}
The total pressure is given by
\begin{equation}
p=\frac 12\dot{\phi}^2-V(\phi ,T).  \label{p}
\end{equation}
The energy-momentum conservation
\begin{equation}
\dot{\rho}+3H(\rho +p)=0,  \label{energymomentum}
\end{equation}
combining with Eq. (\ref{EOMphi}) yields the entropy production equation
\begin{equation}
T\dot{s}+3HTs=\Gamma \dot{\phi}^2.  \label{entropy}
\end{equation}
If the thermal corrections to the potential is little enough, which we'll
see later in the paper is the demand for a workable warm inflationary model,
the radiation energy can be written as $\rho _r=3Ts/4$, the Eq. (\ref
{entropy}) is equivalent to
\begin{equation}
\dot{\rho}_r+4H\rho _r=\Gamma \dot{\phi}^2.  \label{dotrhor}
\end{equation}
Inflation is associated with a slow-roll approximation which consists of
neglecting the highest order terms in the preceding equations, which implies
that the energy is potential dominated and the producing of radiation is
quasi-static. The slow-roll equations in Warm-LQC are:
\begin{equation}
\dot{\phi}=-\frac{V_\phi }{3H(1+R)},  \label{SRdotphi}
\end{equation}
\begin{equation}
Ts=R\dot{\phi}^2,  \label{SRTs}
\end{equation}
\begin{equation}
H^2=\frac{8\pi G}3V(1-z),  \label{SRH}
\end{equation}
\begin{equation}
4H\rho _r=\Gamma \dot{\phi}^2,  \label{SRrho}
\end{equation}
where $z=\rho /\rho _c\simeq V/\rho _c$ in Eq. (\ref{SRH}) is a character
parameter describing the quantum effect in LQC. The validity of the
slow-roll approximation is dependent on a set of slow-roll parameters:
\begin{equation}
\epsilon =\frac{M_p^2}2\left( \frac{V_\phi }V\right) ^2,\ \eta =M_p^2\frac{%
V_{\phi \phi }}V,\ \beta =M_p^2\frac{V_\phi \Gamma _\phi }{V\Gamma },
\label{SRparameter}
\end{equation}
where $M_p^2=1/8\pi G$. There will be two additional parameters describing
the temperature dependence,
\begin{equation}
b=\frac{TV_{\phi T}}{V_\phi },\ c=\frac{T\Gamma _T}\Gamma .  \label{SRbc}
\end{equation}

In warm inflation, the radiation energy density and the universal
temperature has the relationship:
\begin{equation}
\rho _r=C_rT^4.  \label{rhor}
\end{equation}
Considering Eqs. (\ref{Gamma}), (\ref{SRdotphi}), (\ref{SRrho}), and (\ref{rhor}%
), we can get the temperature of the universe during inflation in strong
dissipation regime
\begin{equation}
T=\left( \frac{V_\phi ^2\phi _0^n\tau _0^m}{4HC_r\Gamma_0 \phi ^n}\right) ^{%
\frac 1{4+m}}.  \label{T}
\end{equation}
Either in low or high temperature regime, the criterion for warm inflation $%
T>H$ should de satisfied. When the slow roll parameter $\epsilon \simeq (1+R)%
(1-z)^2/(1-2z)$, $\ddot{a}=0$, which implies the end of the inflation
phase. The number of e-folds in Warm-LQC is given by
\begin{equation}
N(\phi )=\int Hdt=-\frac 1{M_p^2}\int_\phi ^{\phi _e}\frac V{V_\phi }%
(1-z)(1+R)d\phi ^{\prime },  \label{efold}
\end{equation}
the subscript $e$ is used to denote the end of inflation.

\section{\label{sec:level4}Stability Analysis}

To find the conditions for the validity of the slow-roll approximation in the
Warm-LQC scenario, we perform a linear stability analysis to see whether the
system remains close to the slow-roll solution for many Hubble times. The
stability is done around the slow-roll solutions, for we should obtain the
conditions to guarantee they can really act as formal attractor solutions
for the dynamical system. Define a new variable $u=\dot{\phi}$, so that $%
\ddot{\phi}=\dot{u}$. We rewrite Eqs. (\ref{EOMphi}) and (\ref{entropy}) as:
\begin{equation}
\dot{\phi}=u,  \label{dotphi}
\end{equation}
\begin{equation}
\dot{u}=-3H(1+R)u-V_\phi ,  \label{dotu}
\end{equation}
\begin{equation}
T\dot{s}=-3HTs+\Gamma u^2.  \label{dotsT}
\end{equation}
Using the effective Friedmann equation, we obtain the rate of change of the
Hubble parameter is
\begin{equation}
\frac 1H\frac{d\ln H}{dt}=\frac{\dot{H}}{H^2}=-\frac 32\frac{\left(
1-2z\right) \left( u^2+Ts\right) }{(1-z)\left( \frac 12u^2+Ts+V\right) }.
\label{rateHubble}
\end{equation}
Hubble parameter is nearly a constant during slow-roll approximation period,
so we require $|\dot{H}/H^2|\ll 1$. Therefore, we have conditions $Ts\ll V$
and $u^2\ll V$, which are consistent with the inflation realizing
requirement that the energy must be potential dominated. Furthermore, we
should restrict the factor $\frac{1-2z}{1-z}$ is of order of unity, i.e., the
quantum effect shouldn't be too large. When $\rho >\rho _c/2$, the universe
is in the super-inflation phase, the character of this phase is that $\dot{H}%
>0$ and changes fast, but scale factor $a$ varies slowly. The phase is
sometimes called ''fast-roll'' phase and the number of e-folds during that
period is small \cite{superinflation}, so we only focus on the normal
inflation period.

Now we define $\delta \phi $ as a small homogeneous perturbation of the
inflaton field in the slow-roll approximation and the inflaton field can be
expanded as $\phi =\phi _0+\delta \phi $, where $\phi _0$ is the background
part of $\phi $. Similarly, we have $u=u_0+\delta u$ and $s=s_0+\delta s$,
where the subscript `0' denotes the background part of $u$ and $s$. That the
perturbation parts are much less than the background terms is assumed.
The equations for $\phi _0$, $u_0$, and $s_0$ are given by
\begin{equation}
u_0=-\frac{V_{\phi 0}}{3H_0(1+R_0)},  \label{u0}
\end{equation}
\begin{equation}
T_0s_0=R_0u_0^2,  \label{T0s0}
\end{equation}
\begin{equation}
H_0^2=\frac 1{3M_p^2}V_0\left( 1-\frac{V_0}{\rho _c}\right) .  \label{H02}
\end{equation}
Using $s\simeq -V_T$, we have
\begin{equation}
\delta s=-V_{TT}\delta T-V_{\phi T}\delta \phi .  \label{deltas}
\end{equation}
As the condition that the thermal corrections to the potential is
negligible, the relationship $T s_{_T}=3s$ holds, with this condition, we find
\begin{equation}
\delta T=\frac 1{3s_0}\left( T_0\delta s+V_\phi b\delta \phi \right) .
\label{deltaT}
\end{equation}
Similarly, we can calculate the variations of $H$, $\Gamma $, $V_\phi $, and $%
R$ in terms of the fundamental perturbation variables. The results are as
follows:
\begin{equation}
2H_0\delta H=\frac 1{3M_p^2}(1-2z)\left( u_0\delta u+V_\phi \delta \phi
+T_0\delta s\right) ,  \label{deltaH}
\end{equation}
\begin{equation}
\frac{\delta \Gamma }{\Gamma _0}=\left( \frac 1{M_p^2}\frac{V_0}{V_\phi }%
\beta +\frac c{3s_0T_0}V_\phi b\right) \delta \phi +\frac c{3s_0}\delta s,
\label{deltagamma}
\end{equation}
\begin{equation}
\frac{\delta V_\phi }{V_\phi }=\left( \frac 1{M_p^2}\frac{V_0}{V_\phi }\eta +%
\frac{V_\phi }{3s_0T_0}b^2\right) \delta \phi +\frac b{3s_0}\delta s,
\label{deltav}
\end{equation}
\begin{eqnarray}
\frac{\delta R}{R_0} &=&\left( \frac 1{M_p^2}\frac{V_0}{V_\phi }\beta +\frac{%
bcV_\phi }{3s_0T_0}-\frac{1-2z}{2H_0^2}\frac{V_\phi }{3M_p^2}\right) \delta
\phi   \nonumber \\
&&-\frac{1-2z}{2H_0^2}\frac{u_0\delta u}{3M_p^2}+\left( \frac c{3s_0}-\frac{%
1-2z}{2H_0^2}\frac{T_0}{3M_p^2}\right) \delta s.  \label{deltar}
\end{eqnarray}
The equations above will be used repeatedly thereinafter.

For conciseness, we write the equations of small perturbations in a matrix
form
\begin{equation}
\left(
\begin{array}{c}
\delta \dot{\phi} \\
\delta \dot{u} \\
\delta \dot{s}
\end{array}
\right) =E\left(
\begin{array}{c}
\delta \phi  \\
\delta u \\
\delta s
\end{array}
\right) -F,  \label{matrix}
\end{equation}
where $E$ is a $3\times 3$ matrix which can be written:
\begin{equation}
E=\left(
\begin{array}{ccc}
0 & 1 & 0 \\
A & \lambda _1 & B \\
C & D & \lambda _2
\end{array}
\right) .  \label{Ematrix}
\end{equation}
Using Eqs. (\ref{dotphi}) - (\ref{dotsT}) and Eqs. (\ref
{deltaH}) - (\ref{deltar}), we get the elements of the above matrix:
\begin{eqnarray}
A &=&3H_0^2\left[ -\frac{\left( 1+R_0\right) ^2}{R_0}b^2+c\left(
1+R_0\right) b-\frac \eta {1-z}\right.   \nonumber \\
&&\left. +\frac \epsilon {1+R_0}\frac{1-2z}{\left( 1-z\right) ^2}+\frac{R_0}{%
1+R_0}\frac \beta {1-z}\right] ,  \label{A}
\end{eqnarray}
\begin{equation}
B=\frac{H_0T_0}{u_0}\left[ -c+\frac{1+R_0}{R_0}b-\frac \epsilon {\left(
1+R_0\right) ^2}\frac{1-2z}{(1-z)^2}\right] ,  \label{B}
\end{equation}
\begin{eqnarray}
C &=&\frac{3H_0^2u_0}{T_0}\frac{R_0}{1+R_0}\left[ \frac{\left( 1+R_0\right)
^2(1-c)}{R_0}b\right.   \nonumber \\
&&\left. -\beta \frac 1{1-z}+\epsilon \frac{1-2z}{(1-z)^2}\right] ,
\label{C}
\end{eqnarray}
\begin{equation}
D=\frac{H_0u_0}{T_0}\left[ 6R_0-\frac{R_0\epsilon }{\left( 1+R_o\right) ^2}%
\frac{1-2z}{(1-z)^2}\right] ,  \label{D}
\end{equation}
\begin{equation}
\lambda _1=-3H_0(1+R_0)-H_0\frac \epsilon {\left( 1+R_0\right) ^2}\frac{1-2z%
}{(1-z)^2},  \label{lambda1}
\end{equation}
\begin{equation}
\lambda _2=-H_0(4-c)-H_0\frac{R_0\epsilon }{\left( 1+R_o\right) ^2}\frac{1-2z%
}{(1-z)^2}.  \label{lambda2}
\end{equation}
The "forcing term" F is a column matrix which can be expressed as
\begin{equation}
F=\left(
\begin{array}{c}
0 \\
\dot{u}_0 \\
\dot{s}_0
\end{array}
\right) .  \label{Fterm}
\end{equation}
In order to make the slow-roll solution an attractor for Warm-LQC, we need
the matrix $E$ to have negative eigenvalues and the forcing term to be small
enough. First we'll study the forcing term. Taking time derivatives of the
Eq. (\ref{u0}) and (\ref{T0s0}), and after some calculation we can get
\begin{equation}
\dot{u}_0=\frac{BC-A\lambda _2}{\lambda _1\lambda _2-BD}u_0,  \label{dotu0}
\end{equation}
\begin{equation}
\dot{s}_0=\frac{AD-C\lambda _1}{\lambda _1\lambda _2-BD}u_0.  \label{dots0}
\end{equation}
Combining the two equations above with expressions for the matrix elements
we can finally get
\begin{eqnarray}
\frac{\dot{u}_0}{H_0u_0} &=&\frac 1\Delta \left[ -\frac{c R_0+c-4}{1+R_0}%
\epsilon \frac{1-2z}{(1-z)^2}+\frac{\eta (c-4)}{1-z}\right.   \nonumber \\
&&\left. +\frac{4R_0}{1+R_0}\frac \beta {1-z}+3(1+R_0)bc\right] ,
\label{dotu0H0u0}
\end{eqnarray}
\begin{eqnarray}
\frac{\dot{s}_0}{H_0s_0} &=&\frac 3\Delta \left\{ \frac{3+R_0}{1+R_0}%
\epsilon \frac{1-2z}{(1-z)^2}\right.   \nonumber \\
&&-\frac{2\eta }{1-z}+\frac{R_0-1}{R_0+1}\frac \beta {1-z}  \nonumber \\
&&\left. +\left[ (1+R_0)^2+c(R_0^2-1)\right] \frac b{R_0}\right\} .
\label{dots0H0s0}
\end{eqnarray}
Using Eq. (\ref{deltaT}) and the equations above, we can get
\begin{eqnarray}
\frac{\dot{T}_0}{H T_0} &=&\frac 1\Delta \left[ \frac{3+R_0}{1+R_0}\epsilon \frac{%
1-2z}{(1-z)^2}-\frac{2\eta }{1-z}\right.   \nonumber \\
&&\left. +\frac{R_0-1}{R_0+1}\frac \beta {1-z}-\frac{3(1+R_0)^2}{R_0}%
b\right] ,  \label{dotT0H0T0}
\end{eqnarray}
where
\begin{equation}
\Delta =(4+c)R_0+4-c-2(1+R_0)b.  \label{Delta}
\end{equation}
Furthermore, we can get the relationship for $\dot{H}_0/H_0^2$
\begin{equation}
\frac{\dot{H}_0}{H_0^2}=-\frac{1-2z}{(1-z)^2}\frac \epsilon {1+R_0}\ll 1.
\label{dotH0}
\end{equation}
A small forcing term requires $|\dot{u}_0/Hu_0|\ll 1$ and $|\dot{s}%
_0/Hs_0|\ll 1$, and through Eqs. (\ref{dotu0H0u0})-(\ref{dots0H0s0}), and Eqs.(\ref{Delta})-(\ref{dotH0}) we find sufficient conditions for these are
\begin{eqnarray*}
\epsilon  &\ll &(1+R)\frac{(1-z)^2}{1-2z},\quad |\eta |\ll (1+R)(1-z), \\
|\beta | &\ll &(1+R)(1-z),\quad |b|\ll \frac R{1+R}.
\end{eqnarray*}
For the sake of comparison with classical warm inflation and simplicity in later sections,
we redefine the slow-roll parameters in Warm-LQC
\begin{equation}
\tilde{\epsilon}=\epsilon \frac{1-2z}{(1-z)^2},\ \tilde{\eta}=\eta \frac 1{%
1-z},\ \tilde{\beta}=\beta \frac 1{1-z}.  \label{tilde}
\end{equation}
Using the new parameters, the slow-roll conditions in Warm-LQC will have the same form with the
classical case ($\tilde{\epsilon}\ll1+R$, $\tilde{\eta}\ll1+R$ and $\tilde{\beta}\ll1+R$) \cite{Ian2008}. Consistently, in the limit $\rho_c\rightarrow\infty$, naturally we have $%
\tilde{\epsilon}\rightarrow \epsilon $,$~\tilde{\eta}\rightarrow \eta $,$~%
\tilde{\beta}\rightarrow \beta $, i.e. the new defined slow-roll parameters are reduced to the classical ones.

During normal inflation phase in LQC ($0<z<1/2$), $(1-z)^2/\left( 1-2z\right) >1$, so the
condition for $\epsilon $ is less restricted and inflation can last longer
than in the classical case, but the conditions for $\eta $ and $\beta $ are more
restricted due to the quantum correction. Furthermore, the range of slow-roll
parameters $\epsilon $ and $\eta $ are all enlarged by a factor $%
(1+R)$ compared to standard inflation (the slow-roll condition in standard inflations are
$\epsilon\ll1$ and $\eta\ll1$ \cite{Liddle2000,BereraIanRamos}). The slow-roll parameter $b$ is much
more small than others, which indicates a negligible thermal correction to
the potential. The condition for $b$ guarantees the equivalency between Eqs. (%
\ref{entropy}) and (\ref{dotrhor}). With a small $b$, we have
\begin{equation}
\Delta \approx (4+c)R_0+4-c.  \label{delta}
\end{equation}

Now we will find conditions for the matrix having negative or positive but
of order ${\cal O}(\epsilon /R)$ eigenvalues. Through the slow-roll
conditions we have obtained, we find the matrix elements $A$ and $C$ are
very small, so we have the characteristic equation for the matrix $E$ as
\begin{eqnarray}
det(\lambda I-E) &\simeq &\left|
\begin{array}{ccc}
-\lambda  & 1 & 0 \\
0 & \lambda _1-\lambda  & B \\
0 & D & \lambda _2-\lambda
\end{array}
\right|   \nonumber \\
&=&-\lambda (\lambda _1-\lambda )(\lambda _2-\lambda )+BD\lambda   \nonumber
\\
&=&0.  \label{charEq}
\end{eqnarray}
Obviously there is a very small eigenvalue $\lambda \ll \lambda _1,\lambda _2
$:
\begin{equation}
\lambda \simeq \frac{-BC-A\lambda _2}{\lambda _1\lambda _2-BD-A}.
\label{charlambda}
\end{equation}
The other two eigenvalues satisfy
\begin{equation}
\lambda ^2-(\lambda _1+\lambda _2)\lambda +\lambda _1\lambda _2-BD=0.
\label{twoeigen}
\end{equation}
Both eigenvalues are negative if $\lambda _1+\lambda _2<0,~~\lambda
_1\lambda _2-BD>0$ are satisfied. Substituting the expression for $\lambda _1
$, $\lambda _2$, $B$ and $D$ we can finally get
\begin{equation}
|c|<4.  \label{c}
\end{equation}
Now we have obtained all slow-roll conditions in the Warm-LQC scenario. We find
that the conditions for $\epsilon $, $\eta $, and $\beta $ are modified by
the quantum correction compared to warm inflation in classical universe \cite
{Ian2008}, and the conditions for $b$ and $c$ are the same as in \cite
{Ian2008} for they only represent the temperature dependence in inflation.

In the next two sections, we will discuss in the regime where the slow-roll
conditions are satisfied. The slow-roll condition for $\epsilon $ naturally
guarantees the potential dominating by
\begin{equation}
\frac{\rho _r}V=\frac{R\epsilon }{2(1+R)^2(1-z)}\ll 1.  \label{rhorV}
\end{equation}

\section{\label{sec:level5}The case of ${\bf \Gamma =\Gamma (\phi )}$}

In this section, we will focus on the case where the dissipative coefficient
is independent of temperature, i.e.
\begin{equation}
\Gamma =\Gamma (\phi ).
\end{equation}
This is the case for $m=0$ in the Eq. (\ref{Gamma}) and using the definition
of $c$, we can see the slow-roll parameter $c=m=0$ here and the expression for
temperature is reduced to
\begin{equation}
T=\left( \frac{V_\phi ^2\phi _0^n}{4HC_r\Gamma _0\phi ^n}\right) ^{1/4}.
\label{Tm0}
\end{equation}
In the $m=0$ case, the spectrum of the field perturbation amplitude caused
by thermal fluctuations can be written in an analytical form \cite{ChrisIan}
\begin{equation}
P_\phi =\frac{\sqrt{3\pi }}2HT\sqrt{1+R}.  \label{Pphi}
\end{equation}
The complicated
treatment of perturbations of the effective Hamiltonian in LQC in details can be seen in \cite{Bojowald2011,LQCperturbation},
which is beyond the scope of our paper. And we point out here that the cosmological perturbations in LQC have been investigated systematically
in many references such as \cite{Bojowald2011,LQCperturbation,lingyiJCAP}. As stated in \cite{lingyiJCAP,Herrera2010}
, the quantum effect is weak when horizon crossing, which we will see in our later paragraph also. So for simplicity, we will follow the treatment in Ref. \cite{lingyiJCAP,Herrera2010} for LQC, the corresponding curvature perturbation is given by ${\cal R}=\left( H/\dot{\phi}\right) \delta \phi $,
so we can get the amplitude of scalar perturbation as
\begin{equation}
P_s=\left( \frac H{\dot{\phi}}\right) ^2P_\phi =\frac{\sqrt{3\pi }}2\frac{%
H^3T}{\dot{\phi}^2}\sqrt{1+R},  \label{Ps0}
\end{equation}
which is valued at the Hubble radius crossing $k=aH$.  We can see that $\dot\phi$ is in the denominator, and we may ask what will happen if $\dot\phi=0$. $\dot\phi$ may vanish when the loop quantum effect driving the inflaton to the potential hill during the non-slow-roll regime in loop quantum scenario. Fortunately, as stated in \cite{dotphi0}, the expression for scalar perturbation behaves well even when $\dot\phi=0$, and the expression of $P_s$
\begin{equation}\label{Psphi0}
    P_s=\frac{\sqrt{3\pi}}{2}\frac{9H^5T}{V^2_{\phi}}(1+R)^{\frac52}
\end{equation}
is still valid. Furthermore, the validity of slow-roll conditions keep us away from the $\dot\phi=0$ case.
Substituting Eqs. (\ref{SRdotphi}), (\ref
{SRH}) and (\ref{Tm0}) to the equation above yields:
\begin{eqnarray}
P_s &=&\sqrt{\frac{243\pi }8}(1+R)^{2/5}V_\phi ^{-3/2}  \nonumber \\
&&\times \left[ \Gamma _0C_r\left( \frac \phi {\phi _0}\right) ^n\right]
^{-1/4}\left[ \frac{V(1-z)}{3M_p^2}\right] ^{19/8}.  \label{Ps1}
\end{eqnarray}
The spectral index is defined by
\begin{equation}
n_s-1=\frac{d\ln P_s}{d\ln k}.  \label{indexn0}
\end{equation}
Since $H$ varies slowly during the inflation, we have $d\ln k=d(\ln
aH)\simeq d\ln a=Hdt$, and then
\begin{equation}
n_s-1=\frac{\dot{P_s}}{HP_s}.  \label{indexn1}
\end{equation}
By virtue of Eq. (\ref{deltar}) and Eqs. (\ref{dotu0H0u0})-(\ref{dotH0}), we obtain
\begin{eqnarray}
&&n_s-1  \nonumber \\
&=&\frac 1{\Delta }\left[ -\frac{9R-5c+17}{1+Q}\epsilon \frac{1-2z}{%
(1-z)^2}-\frac{9R+1}{1+R}\frac \beta {1-z}\right.   \nonumber \\
&&+\frac{-3cR+6R+6-2c}{1+R}\frac \eta {1-z}  \nonumber \\
&&\left. -\frac{3(1+R)(5cR+2R+2)}{2R}b\right] .  \label{indexn2}
\end{eqnarray}
In the case $c=0$, by using new slow-roll parameter defined in Eq. (\ref
{tilde}) the equation above can reduce to
\begin{eqnarray}
n_s-1 &=&\frac 1{\Delta }\left[ -\frac{9R+17}{1+R}\tilde{\epsilon}+6%
\tilde{\eta}\right.   \nonumber \\
&&\left. -\frac{9R+1}{1+R}\tilde{\beta}-\frac{3(1+R)^2}Rb\right] .
\label{indexn3}
\end{eqnarray}
From the slow-roll conditions we have obtained, we find $n_s\approx 1$, i.e.
we have a nearly scale-invariant power spectrum which is consistent with
observation.

As stated in Sec. \ref{sec:level3}, we have $R\gg1$ in the strong regime of Warm-LQC, thus
\begin{equation}
n_s-1=-\frac 9{4R}\tilde{\epsilon}-\frac 9{4R}\tilde{\beta}+\frac 3{2R}%
\tilde{\eta}-\frac 34b,  \label{indexn4}
\end{equation}
which agrees with a partial result in Ref. \cite{Lisa2004} in form. In the
weak regime of Warm-LQC ($R\ll 1$), thermal fluctuations lead to the
spectral index
\begin{equation}
n_s-1=-\frac{17}4\tilde{\epsilon}+\frac 32\tilde{\eta}-\frac 14\tilde{\beta}-%
\frac 3Rb.  \label{indexn5}
\end{equation}
Although in the weak regime, the spectral index does not approach that in
the standard inflation when $R\rightarrow 0$, the reason we will see in
later section.

Now we will calculate the running of the spectral $\alpha_s =\frac{dn_s}{d\ln k%
}=\frac{dn_s}{Hdt}$. For simplicity, we only focus on the strong regime,
where the running of the spectral can be expressed as
\begin{eqnarray}
\alpha _s &=&\frac 1R\left[ -\frac 94\tilde{\epsilon}^{\prime }-\frac 94%
\tilde{\beta}^{\prime }+\frac 32\tilde{\eta}^{\prime }\right]   \nonumber \\
&&-\frac{R^{\prime }}R\left( n_s-1+\frac 34b\right) -\frac 34b^{\prime },
\label{running}
\end{eqnarray}
where
\begin{equation}
\tilde{\epsilon}^{\prime }=\frac{8z^2\epsilon ^2(2\epsilon -\eta )}{\left(
1-z\right) ^5\left( 1+R\right) ^2},
\end{equation}
\begin{equation}
\tilde{\eta}^{\prime }=\frac{2\epsilon ^2z(\xi -2\eta )}{\left( 1-z\right)
^4\left( 1+R\right) ^2},
\end{equation}
\begin{equation}
\tilde{\beta}^{\prime }=\frac{2\epsilon z\left( 2\epsilon \sigma +\beta \eta
-2\epsilon \beta -2\epsilon \delta \right) }{\left( 1-z\right) ^4\left(
1+R\right) ^2},
\end{equation}
\begin{equation}
R^{\prime }=-\frac{R\beta }{(1+R)(1-z)}+\frac{R(1-2z)\epsilon }{(1+R)\left(
1-z\right) ^2},
\end{equation}
\begin{equation}
b^{\prime }=\frac{b\eta }{(1-z)(1+R)}-\frac \zeta {(1-z)(1+R)}.
\end{equation}
In the equations above, we have $^{\prime }=d/d\ln k$,$~\xi =2M_p^2V_{\phi
\phi \phi }/V$,$~\sigma =M_p^2\Gamma _{\phi \phi }/\Gamma $,$~\delta
=M_p^2\Gamma _\phi ^2/\Gamma ^2$,$~\zeta =M_p^2TV_{\phi \phi T}/V$, and $\xi
$, $\sigma $, $\delta $, $\zeta $ are all small quantities. If we use $%
\lambda $ to refer to slow-roll parameters in general in this section, we
find in strong regime of Warm-LQC, the spectral index is of order ${\cal O}%
(\lambda /R)$, and the running of the spectral is of order ${\cal O}(\lambda
^2/R^2)$ (the dominated term is the second one in the expression of $\alpha_s$ and $\alpha_s$ is a little smaller by the quantum correction).
As $R\gg 1$, we probably get a nearly scale-invariant spectrum.

Tensor perturbations do not couple strongly to the thermal background, and
so gravitational waves are only generated by the quantum fluctuations, as in
standard inflation \cite{Taylor2000},
\begin{equation}
P_T=\frac 2{M_p^2}\left( \frac H{2\pi }\right) ^2=\frac{V(1-z)}{6\pi ^2M_p^4}%
\sim \frac V{M_p^4}.  \label{tensor}
\end{equation}
Amusingly, we can see from the equation above that the gravitational wave
spectrum and the quantum effect ($z\simeq V/\rho _c$) have the same order.
If the primordial tensor perturbations are seen in the CMB sky, so does the
quantum effect.

The spectral index of tensor perturbation is
\begin{equation}
n_T=\frac{\dot{P_T}}{HP_T}=-\frac{2\tilde{\epsilon}}{1+R},
\label{tensorindex}
\end{equation}
and the tensor-to-scalar ratio is
\begin{equation}
r=\frac{P_T}{P_s}=\frac{2H\epsilon }{\sqrt{3}\pi ^{5/2}\left( 1+R\right)
^{5/2}T\left( 1-z\right) ^2},  \label{ratio}
\end{equation}
which is much smaller than in standard inflation due to the factor $%
(1+R)^{5/2}$ in the denominator.

Through the analysis above, we can find that both scalar and tensor perturbations are depressed by
loop quantum correction to varying degrees compared to the classical case. We can get a nearly scale-invariant power spectrum for warm inflation in both classical and loop cosmology. The expression for spectral index $n_s$ and $n_T$ are
modified by quantum correction but has the similar form with those in classical universe \cite{Ian2008}. But the expression for the running of spectral $\alpha_s$ is very different from the classical one \cite{ZhangYi2009}, the loop
correction can result in some additional terms but doesn't change the magnitude of the running of spectral. Furthermore, the tensor-to-scalar ratio is a little larger by a factor $1/(1-z)^{3/2}$ than in classical warm inflation.

Now we will study the amplitude of quantum effect and see whether it can be
observed. The expected amplitude of the quantum effect can be characterized by a
parameter $z$ which appears in the power spectrum of Warm-LQC, and we can
determine its value by fitting the theoretical results to observational
data. The WMAP observation results give the values for scalar power spectrum
$P_s\simeq 2.3\times 10^{-9}$ and the tensor-to-scalar ratio $r<0.2$, and by
using Eq. (\ref{tensor}) we can obtain $z=\frac V{\rho _c}<\frac{0.2}{0.41}%
\frac{6\pi ^2}{1-z}P_s$ is a small quantity when horizon crossing, so we can
use the condition to simplify the calculation in the following. Using Eq. (%
\ref{Psphi0}) we get
\begin{eqnarray}
z &=&\frac {P_s^2\tilde{\epsilon}^2}{4\times 0.41\pi ^3(1+R)^5}\left( \frac{%
M_p}T\right) ^2  \nonumber \\
&\simeq &10^{-18}\frac{\tilde{\epsilon}^2}{(1+R)^5}\left( \frac{M_p}T\right)
^2,  \label{z0}
\end{eqnarray}
which is fairly depressed by the factor $10^{-18}$ and $(1+R)^5$ in the
denominator. As in standard inflation in LQC \cite{lingyiJCAP} and warm inflation in LQC \cite{Herrera2010}, we also
come to the conclusion that the LQC quantum effect is too tiny to be observed today. For $R\gg 1$, we have
\begin{equation}
z\ll 10^{-18}\tilde{\epsilon}^2\left( \frac{M_p}T\right) ^2.  \label{z1}
\end{equation}
The upper limit for $z$ is dependent on temperature and $\tilde{\epsilon}$ and the
upper limit increases when temperature decreases.

\subsection*{\label{sec:level51}A separable potential example}

Let us consider a separable potential in SUSY case \cite{Campo2010}:
\begin{equation}
V(\phi ,T)=v_1(T)+v_2(\phi ),  \label{separablepotential}
\end{equation}
where
\begin{equation}
v_1(T)=-\frac{\pi ^2}{90}g_{*}T^4-\frac 1{12}m_\phi ^2T^2,  \label{v1T}
\end{equation}
\begin{equation}
v_2(\phi )=\frac 12m_\phi ^2\phi ^2.  \label{v2phi}
\end{equation}
SUSY is a good mechanism which can suppress the thermal corrections to the
potential, so in this case, the potential can be written in a separable
form. The slow-roll parameter $b=0$ here can certainly satisfy the
slow-roll conditions and the other slow-roll parameters are given by
\begin{equation}
\epsilon =\frac{M_p^2m_\phi ^4\phi ^4}{2V^2},\ \eta =\frac{M_p^2m_\phi ^2}V%
,\ \beta =\frac{nM_p^2m_\phi ^2}V.  \label{threeSR}
\end{equation}
Because of Eq. (\ref{rhorV}), we have $V\simeq v_2$, so the three parameters
above can be written as
\begin{equation}
\epsilon \simeq \frac{2M_p^2}{\phi ^2},\ \eta \simeq \frac{2M_p^2}{\phi ^2}%
,\ \beta \simeq \frac{2nM_p^2}{\phi ^2}.  \label{threeSR1}
\end{equation}
With this potential, the slow-roll conditions hold in the regime $\phi \gg
\frac{M_p}{\sqrt{1+R}}$. If the $R$ is big enough, the lower bound can be
much smaller than Planck scale, while in the standard inflation case with a
chaotic potential, the range of the inflaton is $\sqrt{2}M_p<\phi <\sqrt{4N+2%
}M_p$ ($N$ is the number of e-folds), which is too big to build a consistent
inflationary model \cite{Berera2005,BereraIanRamos}. Many kinds of monomial
potential in standard inflation suffer the problem of having an overlarge
inflaton amplitude, which can be eliminated in a warm inflation scenario.
Hence the warm inflation can take place for a wider range of potentials than
standard inflation as the slow-roll conditions imply.

Using $\rho _r=3Ts/4$ and the expression for the separable potential, we
have $C_r=\frac{\pi ^2}{30}g_{*}$ ($g_{*}$ is a constant and $g_{*}\approx 100$ for the radiation field) in Eq. (\ref{rhor}) and the
strength of dissipation parameter is given by
\begin{equation}
R=\frac{M_p\Gamma _0(\phi /\phi _0)^n}{\sqrt{\frac 32(1-z)}m_\phi \phi }.
\label{dissipationR}
\end{equation}
By using $V_\phi =m_\phi ^2\phi $ and Eq. (\ref{Tm0}) we obtain the scalar
power spectrum in a concrete form
\begin{eqnarray}
P_s &=&\sqrt{\frac{243\pi }8}\left( \frac 1 3\right)
^{19/8}(1+R)^{5/2}(1-z)^{19/8}  \nonumber \\
&&\times C_r^{-1/4}\left( \frac{\phi _0}\phi \right) ^{n/4}\left( \frac{M_p}{m_\phi }%
\right) ^3\left( \frac{M_p}\phi \right) ^{3/2}  \nonumber \\
&&\times \left( \frac{M_p}{\Gamma _0}\right) ^{1/4}\left( \frac V{M_p^4}%
\right) ^{19/8}.  \label{Ps2}
\end{eqnarray}
The last factor must be very small to obtain an observation consistent $P_s$
with many factors ahead are big, so the quantum effect characteristic
parameter is a tiny one. By using Eq. (\ref{efold}) we can estimate the
amplitude of inflaton when horizon crossing $k=aH$ (denoted by $\phi _{*}$),
$\phi _{*}\sim \sqrt{(4N+2)/(1+R)}M_p$. In strong dissipation regime, we
take $R\simeq 10^2$, $g_{*}\sim 100$, $\Gamma _0\simeq 10^{-6}M_p$, $\phi
_0\sim M_p$ \cite{Campo2010}, then we have $\phi _{*}\sim M_p$. Setting $P_s$
to the observed value $\approx 2\times 10^{-9}$ and using the parameters
fixed above leads to $m_\phi \simeq 10^{-8}M_p$.

The spectral index is given by
\begin{eqnarray}
n_s-1 &=&\frac 1{1+R}\frac{M_p^2}{2\phi ^2}\left[ -\frac{9R+17}{1+R}\frac{%
1-2z}{(1-z)^2}\right.   \nonumber \\
&&\left. +\frac 6{1-z}-\frac{9R+1}{1+R}\frac n{1-z}\right] ,  \label{indexn6}
\end{eqnarray}
which is of order ${\cal O}(M_p^2/R\phi ^2)$, and a nearly scale-invariant
power spectrum is guaranteed by slow-roll conditions.

In the strong dissipation regime:
\begin{equation}
n_s-1=\frac{-3}{2R(1-z)}\frac{M_p^2}{\phi ^2}(3n+1).  \label{indexn7}
\end{equation}
When $n>-1/3$, the spectrum is red, and when $n<-1/3$, the spectrum can be
blue. The power spectrum of tensor perturbation is
\begin{equation}
P_T=\frac{V(\phi ,T)(1-z)}{6\pi ^2M_p^4}\simeq \frac{m_\phi ^2\phi
_{*}^2(1-z)}{12\pi ^2M_p^4}.  \label{Pt0}
\end{equation}
We can see from the equation above that temperature may leave its imprint on
tensor perturbation in Warm-LQC scenario, while in standard inflation tensor
perturbation is zero-temperature. This effect should be very tiny since
tensor perturbation itself is small.

The number of e-folds is given by
\begin{eqnarray}
N &=&-\frac 1{M_p^2}\int_{\phi _{*}}^{\phi _e}\frac \phi 2\left( 1-\frac{%
m_\phi ^2\phi ^2}{2\rho _c}\right) (1+R)d\phi ^{\prime }  \nonumber
\label{efolds} \\
&\approx &\frac{(1+R)\phi _{*}^2}{4M_p^2},
\end{eqnarray}
where we use the conditions that $z$ is small after horizon crossing and $%
\phi _{*}\gg \phi _e$ ($\phi _e$ is the amplitude of inflaton at the
end of inflation $\phi _e$ $\approx M_p/\sqrt{1+R}$).

\section{\label{sec:level6}The case of ${\bf \Gamma =\Gamma (\phi ,T)}$}

In this section, we will focus on the case where the dissipative coefficient
is dependent of temperature, i.e.
\begin{equation}
\Gamma=\Gamma(\phi,T).
\end{equation}
This is the case for $m\neq0$ in the Eq. (\ref{Gamma}) and by definition $%
c=m $. The temperature is given by Eq. (\ref{T}).

In this case the fluctuation power spectrum $P_\phi $ is determined by two
coupled equations, and it's hardly to get analytic results \cite{Lisa2004}. Previously most work about
warm inflation focuses on the power spectrum of the temperature independent case shown by Eq. (\ref{Ps0}).
In Ref. \cite{ChrisIan}, the authors pay attention to the temperature dependent case and give the
analytic approximation solution and numerical simulation. Here we take the
new form of approximation solution for the power spectrum proposed in Ref. \cite
{ChrisIan} which should be used in the $\Gamma _T\neq 0$ case:
\begin{equation}
P_s=\left( \frac H{\dot{\phi}}\right) ^2P_\phi =\frac{\sqrt{3\pi }}2\frac{%
H^3T}{\dot{\phi}^2}\sqrt{1+R}\left( 1+\frac R{r_c}\right) ^{3c},  \label{Ps3}
\end{equation}
where $r_c$ varies slowly in the range $0\leq c\leq 3$, for example $%
r_1\approx 8.53$, $r_2\approx 7.66$, $r_3\approx 7.27$. The formula is
consistent with the old one when $c=m=0$. The $\dot\phi=0$ case doesn't affect our analysis as stated in the previous chapter. After some cockamamie calculation
we can obtain the spectral index for the new scalar power spectrum:
\begin{eqnarray}
&&n_s-1  \nonumber \\
&=&\frac 1{\Delta }\left\{ \left[ \frac{-17+5m-9R}{1+R}+\frac{6mR(m+2)}{%
1+R/r_c}\right] \tilde{\epsilon}\right.   \nonumber \\
&&+\left( \frac{-3mR+6R-2m+6}{1+R}-\frac{6m^2R}{1+R/r_c}\right) \tilde{\eta}
\nonumber \\
&&-\left( \frac{9R+1}{1+R}+\frac{12mR}{1+R/r_c}\right) \tilde{\beta}
\nonumber \\
&&\left. -\left[ \frac{3\left( 5mR+2R+2\right) }{2R}+\frac{9m^2(1+R)}{1+R/r_c%
}\right] (1+R)b\right\} . \nonumber \\
\label{indexn8}
\end{eqnarray}
Many new terms are added compared to Eq. (\ref{indexn3}) but we still have $%
n_s\approx 1$ as the slow-roll conditions hold. We can express the additional
terms as
\begin{eqnarray}
\triangle n_s &=&\frac 1{\Delta (1+R/r_c)}\left[ 6mR(m+2)\tilde{\epsilon}%
-6m^2R\tilde{\eta}\right.   \nonumber \\
&&\left. -12mR\tilde{\beta}-9m^2(1+R)^2b\right] +\frac 1{\Delta }\left[
\frac{5m}{1+R}\tilde{\epsilon}\right.   \nonumber \\
&&\left. -\frac{3mR+2m}{1+R}\tilde{\eta}-\frac{15m(1+R)}2\right] ,
\end{eqnarray}
when $m\rightarrow 0$, $\triangle n_s\rightarrow 0$.

In the strong dissipation regime:
\begin{eqnarray}
n_s-1 &=&\frac 1{(4+m)R}\left\{ \left[ -9+6r_cm(m+2)\right] \tilde{\epsilon}%
\right.   \nonumber \\
&&+\left( -3m+6-6r_cm^2\right) \tilde{\eta}-(9+12r_cm)\tilde{\beta}
\nonumber \\
&&\left. -\left[ \frac 32(5mR+2R)-9r_cm^2R\right] b\right] ,  \label{indexn9}
\end{eqnarray}
\begin{eqnarray}
\triangle n_s &=&\frac m{(4+m)R}\left[ 6r_c(m+2)\tilde{\epsilon}-(6r_cm-3)%
\tilde{\eta}\right.   \nonumber \\
&&\left. -12r_c\tilde{\beta}-\left( 9r_cmR+\frac{15}2R\right) b\right] ,
\label{deltaindex1}
\end{eqnarray}
where we can see that $\triangle n_s$ is of order ${\cal O}(\lambda r_c/R)$.
With the new form of the power spectrum we still have a nearly
scale-invariant spectrum but the departure from 1 for spectral index is
slightly bigger than the old one. In this section, we refer "old" as the
results obtained by using the power spectrum given by Eq. (\ref{Ps0}).

In the weak dissipation regime:
\begin{eqnarray}
n_s-1 &=&\frac 1{4-m}\left[ (-17+5m)\tilde{\epsilon}-\tilde{\beta}\right.
\nonumber \\
&&\left. +(6-2m)\tilde{\eta}-(\frac 3R+9m^2)b\right] ,  \label{indexn10}
\end{eqnarray}
\begin{equation}
\triangle n_s=\frac m{4-m}\left( 5\tilde{\epsilon}-2\tilde{\eta}-\frac{15}2%
b-9bm\right) .  \label{deltaindex2}
\end{equation}
The running of the spectral in strong dissipation regime is given by
\begin{eqnarray}
\alpha _s &=&\frac 1{(4+m)R}\left\{ \left[ (-9+6r_cm(m+2)\tilde{\epsilon}%
^{\prime }\right. \right.   \nonumber \\
&&+(-3m+6-6r_cm^2)\tilde{\eta}^{\prime}-(9+12r_cm)\tilde{\beta}^{\prime }  \nonumber
\\
&&\left. -\left( \frac 32(5m+2)+9r_cm^2\right) R\right] b^{\prime }
\nonumber \\
&&\left. -\left[ \frac 32(5m+2)+9r_cm^2\right] bR^{\prime }\right\}
\nonumber \\
&&-(n_s-1)\frac{R^{\prime }}R,  \label{running1}
\end{eqnarray}
where the parameters are already given in the preceding part of the paper.
The running is still of order ${\cal O}(\lambda ^2/R^2)$.

Tensor perturbation is not coupled to the radiation, so the tensor power
spectrum does not change. The tensor-to-scalar ratio is given by
\begin{equation}
r=\frac{P_T}{P_s}=\frac{2H\epsilon \left( 1+R/r_c\right) ^{-3m}}{\sqrt{3}\pi
^{5/2}(1+R)^{5/2}T(1-z)^2},  \label{ratio1}
\end{equation}
which is smaller than the old one if $m>0$.

There are also some differences between warm inflation in classical and loop cosmology, and most are the same as we stated in the Sec. \ref{sec:level5}. So we don't state again here to avoid repetition.

The quantum characteristic
parameter $z$ has the amplitude constrained by WMAP observations
\begin{eqnarray}
z &\simeq &10^{-18}\frac{\tilde{\epsilon}^2}{(1+R)^5}\left( \frac{M_p}T%
\right) ^2\left( 1+\frac R{r_c}\right) ^{-6m}  \nonumber \\
&\ll &10^{-18}\tilde{\epsilon}^2\left( \frac{M_p}T\right) ^2\left( 1+\frac R{%
r_c}\right) ^{-6m},
\end{eqnarray}
which is depressed by a factor $(1+R/r_c)^{-6m}$ if $m>0$, so the energy
scale is lower than the old case.

From the above analysis, we reach the conclusion that for $m>0$, the new
form of the power spectrum gives a slightly bigger spectral index, a lower
tensor-to scalar ratio and the lower energy scale when horizon crossing.

\subsection*{\label{sec:level61}A separable potential example}

In order to compare to the $m=0$ case, here we also take the potential in
Eq. (\ref{separablepotential}) in SUSY background. The slow-roll parameters
are all the same as in the old case, we don't list them here to avoid
repetition and the expression for the strength of dissipation parameter is
\begin{equation}
R=\frac{M_p\Gamma _0(\phi /\phi _0)^n(T/\tau _0)^m}{\sqrt{\frac 32(1-z)}%
m_\phi \phi }.  \label{R1}
\end{equation}
The scalar power spectrum in a concrete form is given by
\begin{eqnarray}
P_s &=&\frac{9\sqrt{3\pi }}2\left( \frac 14\right) ^{\frac 1{4+m}}\left(
\frac 16\right) ^{\frac{5m+19}{2(4+m)}}\left( 1+R\right) ^{5/2} C_r^{-1/(4+m)} \nonumber \\
&&\times \left( 1+\frac R{r_c}\right) ^{3m}\left( 1-z\right) ^{\frac{5m+19}{%
2(4+m)}}\left( \frac{\phi _0}\phi \right) ^{\frac n{4+m}}\left( \frac{m_\phi }{%
M_p}\right) ^{\frac{m+7}{m+4}}  \nonumber \\
&&\times \left( \frac \phi {M_p}\right) ^{\frac{3m+13}{4+m}}\left( \frac{%
\tau _0}{M_p}\right) ^{\frac m{4+m}}\left( \frac{M_p}{\Gamma _0}\right) ^{%
\frac 1{4+m}}  \label{Ps4}
\end{eqnarray}
For example,when $m=3$, using the parameters fixed at the same value as in
the previous section and setting $P_s$ to the observed value $\approx
2\times 10^{-9}$ finally lead to $m_\phi \simeq 10^{-14}M_p$. The constraint
is stronger than $m=0$ case ($m_\phi =10^{-8}M_p$), and even when $m=0$ in
Warm-LQC, the constraint is stronger than that in standard inflation where $%
m_\phi \simeq 10^{-7}M_p$ \cite{BereraIanRamos}. We can conclude that for the
chaotic potential case, Warm-LQC with $\Gamma \propto T^m$ ($m>0$) requires a
smaller inflaton mass, which is consistent with our previous analysis that
the new power spectrum with $m>0$ depresses the energy scale of inflation.

The spectral index is
\begin{eqnarray}
&&n_s-1  \nonumber \\
&=&\frac{2M_p^2}{\Delta \phi ^2}\left[ \left( \frac{-17-9R+5m}{1+R}+\frac{%
6mR(m+2)}{1+R/r_c}\right) \frac{1-2z}{\left( 1-z\right) ^2}\right.
\nonumber \\
&&+\left( \frac{-3mR+6R-2m+6}{1+R}-\frac{6m^2R}{1+R/r_c}\right) \frac 1{1-z}
\nonumber \\
&&\left. -\left( \frac{9R+1}{1+R}+\frac{12mR}{1+R/r_c}\right) \frac n{1-z}%
\right] ,  \label{indexn11}
\end{eqnarray}
where the part in the $[~]$ are of order of unity, with slow-roll conditions
hold, the factor $2M_p^2/\Delta \phi ^2\ll 1$, so we have $n_s\approx 1$.

In strong dissipation regime:
\begin{eqnarray}
n_s-1 &=&\frac{6M_p^2}{(4+m)R\phi ^2(1-z)}  \nonumber \\
&&\times \left[ 4r_cm(1-n)-m-3n-1\right] .  \label{indexn12}
\end{eqnarray}
when $m=3$, the spectrum can be red when $n>1$ and blue when $n<1$.

The condition for warm inflation is $T>H$. We now analyze whether the ratio $%
T/H$ will be larger in $m>0$ case. We rewrite the scalar perturbation
Eq. (\ref{Ps3}) as
\begin{equation}
P_s\approx \frac{T^4}{u^2}\frac{H^3}{T^3}(1+R)^{1/2}\left( 1+\frac R{r_c}%
\right) ^{3m}.  \label{Ps5}
\end{equation}
Using Eq. (\ref{T0s0}) and the potential expression Eq. (\ref
{separablepotential}) we can obtain:
\begin{equation}
\frac TH\approx \left( \frac{45}{2\pi ^2g_{*}}\right) ^{1/3}\left(
1+R\right) ^{1/6}\left( 1+\frac R{r_c}\right) ^m\left( \frac R{P_s}\right)
^{1/3}.  \label{T/H}
\end{equation}
The ratio $T/H$ is larger by a factor $\left( 1+\frac R{r_c}\right) ^m$,
when $m>0$, so the thermal effect is more remarkable when dissipative
coefficient has the form $\Gamma \propto T^m$ with $m>0$. And $m<0$ case
will have the opposite behavior. Since $m<0$ corresponds to a non-SUSY case,
we have paid little attention to that case.

The condition for warm inflation $T>H$ can be certainly satisfied by
\begin{equation}
R>g_{*}P_s,
\end{equation}
where $g_{*}$ is of order ${\cal O}(10^2)$, and $P_s$ is of order ${\cal O}%
(10^{-9})$, so very weak dissipation can result warm inflation. Although in the
weak dissipation regime, Warm-LQC results are still different from standard
inflation in LQC such as Eq. (\ref{indexn5}).

\section{\label{sec:level7}Conclusion}

In this paper, we have investigated the warm inflationary scenario in LQC.
We use the general form of dissipative coefficient $\Gamma =\Gamma _0\left(
\phi /\phi _0\right) ^n\left( T/\tau _0\right) ^m$ to study Warm-LQC for the
first time. We give a short review of the effective theory of LQC and the
framework of Warm-LQC. In the scenario of Warm-LQC, the universe has the
temperature shown by Eq. (\ref{T}). We consider the most general and
reasonable case that the potential and dissipative coefficient are
temperature dependent, so we introduced five slow-roll parameters in
Warm-LQC.

We perform a linear stability analysis to determine the conditions
that are sufficient for the system to keep the slow-roll solution as an
attractor. We have proved that the consistency of the slow-roll Warm-LQC
requires the slow-roll parameters satisfy: $\tilde{\epsilon}\ll 1+R$, $|%
\tilde{\eta}|\ll 1+R$, $|\tilde{\beta}|\ll 1+R$ and $|b|\ll R/(1+R)$, $|c|<4$%
. The first three parameters ($\tilde{\epsilon}$, $\tilde{\eta}$, $\tilde{%
\beta}$) are new defined in Warm-LQC, the slow-roll conditions given by them
imply the slow-roll inflation couldn't happen in the super-inflation phase.
The condition on $b$ implies Warm-LQC is only possible when a mechanism such
as SUSY, suppresses thermal corrections to the potential. The conditions on $%
b$ and $c$ are proved to be the same as in classical warm inflation for they
only describe the temperature dependence.

We study two cases when the
dissipative coefficient is independent ($m=0$) and dependent ($m\neq 0$) on
temperature. We use different expressions of power spectrum to investigate
the two cases and obtain their spectral index and the running of spectral
that are different in form. The power spectrum in the temperature independent case we used is the same as \cite{Herrera2010}, which also treats the warm inflationary models in loop quantum cosmology. And we don't treat perturbations of the effective Hamiltonian in loop cosmology in detail. We give a separable potential example which
satisfies the negligible thermal correction condition for both cases. By
using the new power spectrum for the $m\neq 0$ case, we find that the
perturbation amplitude is enhanced when $m>0$ and the contribution of
thermal fluctuations to density fluctuations is more outstanding. And the
temperature dependence for $m>0$ depresses the energy scale of inflation when horizon crossing compared to models with $m=0$. In both cases we have $n_s\approx 1$, but a
bigger spectral index and a smaller tensor-to scalar ratio is obtained in $%
m>0$ case. We also find the ratio $T/H$ is enhanced when $m>0$, so the
thermal effect is more obvious. The differences
between warm inflation in classical and loop university can be seen from the expression
for the power spectrum and spectral index, etc. They all acquire modifications described by the quantum parameter $z$. Among the modifications, the most significant one is that the tensor-to-scalar ratio is larger by $1/(1-z)^{3/2}$. And the running of the spectral has a more complicated expression and is a little smaller than the one in classical universe.
We investigated the quantum effect
characterized by a parameter $z$ in both cases, although quantum effect is
dominated in very early universe, but it leaves tiny imprint on the CMB sky
and can be hardly observed today.

We should note that there are some other properties of this model that are
not considered in this paper but deserve further study. For example, the
non-Gaussian effects during warm inflation \cite{Ian2007} is an important
signature which should be analyzed in Warm-LQC scenario. Furthermore, during
the warm inflation phase, the universe is a multi-component system and the
nonadiabatic entropy perturbation should de present \cite{Wands2000} and
can sometimes leave behind an impression on the curvature fluctuations \cite
{Ian2008}. These should be considered in Warm-LQC scenario and can be our
future work.

\acknowledgments This work was supported by the National Natural Science Foundation of China (Grant Nos. 11175019 and 11235003).

%----------------------------------------------------------------------------------------

\end{document}